\documentclass[aps,prl,superscriptaddress,nofootinbib,twocolumn]{revtex4-1}
\usepackage{amsmath,graphicx}
\usepackage{color}\usepackage{xcolor}

\begin{document}

\title{Primordial fluctuations in heavy-ion collisions}

\author{Fran\c cois Gelis}
\affiliation{
Institut de physique th\'eorique, Universit\'e Paris Saclay, CNRS, CEA, F-91191 Gif-sur-Yvette, France} 
\author{Giuliano Giacalone}
\affiliation{
Institut de physique th\'eorique, Universit\'e Paris Saclay, CNRS, CEA, F-91191 Gif-sur-Yvette, France} 
\author{Pablo Guerrero-Rodr\'iguez}
\affiliation{CAFPE and Departamento de F\'isica Te\'orica y del Cosmos, Universidad de Granada, E-18071 Campus de Fuentenueva, Granada, Spain}
\affiliation{CPHT, CNRS, Institut Polytechnique de Paris, France}
\author{Cyrille Marquet}
\affiliation{CPHT, CNRS, Institut Polytechnique de Paris, France}
\author{Jean-Yves Ollitrault}
\affiliation{
Institut de physique th\'eorique, Universit\'e Paris Saclay, CNRS, CEA, F-91191 Gif-sur-Yvette, France} 

\begin{abstract}
  We present a simple description of the energy density profile created in a nucleus-nucleus collision, motivated by high-energy QCD. 
  The energy density is modeled as the sum of contributions coming from elementary collisions between localized charges and a smooth nucleus. 
  Each of these interactions creates a sharply-peaked source of energy density falling off at large distances like $1/r^2$, corresponding to the two-dimensional Coulomb field of a point charge. 
  Our model reproduces the one-point and two-point functions of the energy density field calculated in the framework of the color glass condensate effective theory, to leading logarithmic accuracy.
  We apply it to the description of eccentricity fluctuations.
  Unlike other existing models of initial conditions for heavy-ion collisions, it allows us to reproduce simultaneously the centrality dependence of elliptic and triangular flow. 
\end{abstract}
\date{\today}

\maketitle
In an ultrarelativistic nucleus-nucleus collision, the strong interaction deposits energy between two crossing nuclei right after the collision takes place.
The resulting profile of energy density is a crucial quantity, as it determines the bulk of particle production.
After a short pre-equilibrium phase~\cite{Gelis:2013rba,Kurkela:2018wud}, this energy density provides the initial condition for the equations of viscous hydrodynamics which govern the subsequent evolution of the fluid~\cite{Romatschke:2017ejr,Gale:2013da}, until it freezes out into individual hadrons which decay~\cite{Mazeliauskas:2018irt} into the detected stable hadrons. 

At ultrarelativistic energies, the energy density can be determined from first principles in the Color Glass Condensate (CGC) approach~\cite{Krasnitz:1999wc,Lappi:2006hq,Albacete:2018bbv}.
It is an interesting case where weakly coupled QCD can be used as an input to model non-perturbative phenomena, such as collective flow~\cite{Giacalone:2019kgg}. 
Event-by-event fluctuations of the energy density are essential for phenomenology~\cite{Alver:2010gr}.
Their magnitude and shape is characterized by the two-point function of the density field, which has recently been calculated analytically in the CGC approach~\cite{Albacete:2018bbv}.

\begin{figure*}[t]
    \centering
    \includegraphics[width=\linewidth]{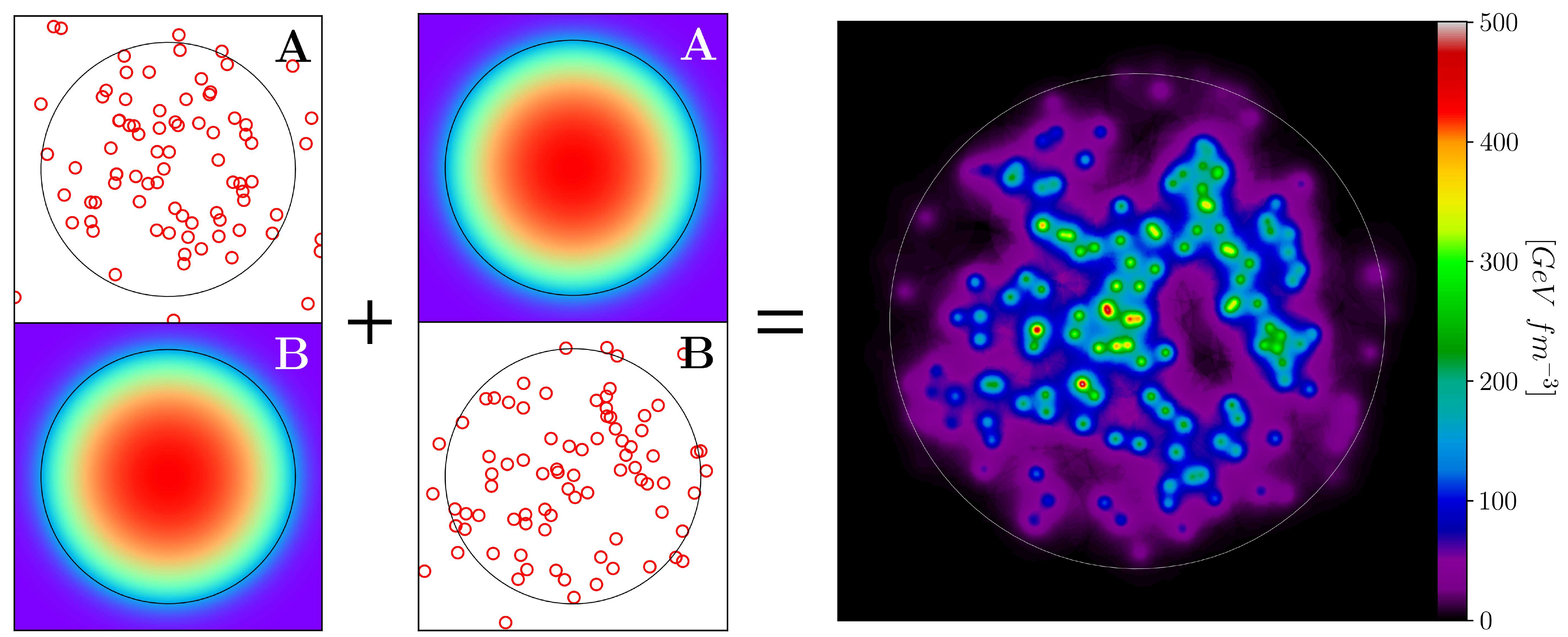}
    \caption{The four plots on the left illustrate the decomposition of the energy density according to Eq.~(\ref{density}) in a central Pb+Pb collision at $\sqrt{s_{\rm NN}}=5.02$~TeV. 
      The upper left and lower right plots depict the positions of sources in nuclei $A$ and $B$ in one event, sampled according to the probability density Eq.~(\ref{densityofsources}), where the saturation momentum at the center of the nucleus is $Q_{s0}=1.24$~GeV, and the infrared cutoff is $m=0.14$~GeV~\cite{Giacalone:2019kgg}.
      The lower left and upper right plots depict the map of saturation momentum over the nuclear area, which is simply proportional to the thickness of the $^{208}$Pb nucleus. 
      The large circles have radius equal to $R=6.62$~fm, i.e., the radius parameter used in the Fermi parametrization of the $^{208}$Pb. The larger plot on the right depicts the energy density, Eq.~(\ref{density}), for that event, obtained upon adding the contributions from sources from $A$ and $B$. We have chosen $g^2=\pi$, corresponding to $\alpha_s=0.25$. With this value, the mean energy density at the center, defined by Eq.~(\ref{1pointcgc}), is $131$~GeV/fm$^3$.}
    \label{fig:1}
\end{figure*}

In this article, we propose a simple model of event-by-event fluctuations of energy density which reproduces the CGC results to leading logarithmic accuracy.
We model the energy density field as the sum of contributions of elementary collisions between a localized color charge and a dense nucleus.
Each elementary collision yields a source of energy density which is independent of rapidity and decreases with transverse distance $r$ like $1/r^2$, characteristic of a 2-dimensional Coulomb field.
We apply our model to the description of elliptic and triangular flows in Pb+Pb collisions.

We denote by $\rho({\bf r})$ the energy density at a transverse point ${\bf r}$ in a single event.
Its value averaged over many events, denoted by $\langle\rho({\bf r})\rangle$, has been calculated in the CGC\footnote{The expressions given in Eq.~(\ref{1pointcgc}) and Eq.~(\ref{xicgc}) were obtained in Ref.~\cite{Albacete:2018bbv} using the McLerran-Venugopalan model, though neglecting the logarithms inherent to the saturation scales in that model.}~\cite{Lappi:2006hq,Albacete:2018bbv}:
\begin{equation}
  \label{1pointcgc}
\langle\rho({\bf r})\rangle=\frac{N_c^2-1}{2g^2N_c}Q_B^2Q_A^2.
\end{equation}
In this equation, $Q_A$ and $Q_B$ are the saturation momenta of the colliding nuclei $A$ and $B$, $N_c$ is the number of colors ($N_c=3$ for QCD), and $g$ denotes the dimensionless coupling constant of QCD.
For now, we treat nuclei as uniform objects with infinite transverse extension, so that the saturation momenta, as well as $\langle\rho({\bf r})\rangle$, are independent of ${\bf r}$. 

The fluctuation around the average value is quantified by the two-point function of the field, defined by:
\begin{equation}
 \label{2point}
S({\bf r}_1,{\bf r}_2)  \equiv \langle\rho({\bf r}_1)\rho({\bf r}_2)\rangle-\langle\rho({\bf r}_1)\rangle\langle\rho({\bf r}_2)\rangle .
\end{equation}
The details of the structure of this correlator as a function of the relative distance $|{\bf r}_1-{\bf r}_2|$ are complicated~\cite{Albacete:2018bbv}, but most of these details are irrelevant for our purposes. 
Indeed, the phenomenology of collective flow in heavy-ion collisions is driven by hydrodynamics, which is a description of the large-scale structure of the system~\cite{Baier:2007ix}.
The relevant fluctuations are not the fluctuations of the energy density itself, but those of its integral over a transverse area much larger than $Q_{A,B}^{-2}$~\cite{Teaney:2010vd,Blaizot:2014nia}, which is the scale of short-range dynamics.
The variance of this integrated fluctuation is proportional to the integral of $S({\bf r}_1,{\bf r}_2)$ over the relative position ${\bf r}_1-{\bf r}_2$~\cite{Giacalone:2019kgg}, which is the most important quantity for phenomenology:
\begin{equation}
 \label{defxi}
 \xi({\bf r})\equiv\int_{\bf s}S\left({\bf r}+\frac{\bf s}{2},{\bf r}-\frac{\bf s}{2}\right),
\end{equation}
where we use the short hand $\int_{{\bf s}}=\int {\rm d}x{\rm d}y$ for the integration over the transverse  plane.
In the CGC, the integrand decreases at large distance like $1/|{\bf s}|^2$, so that the integral diverges logarithmically at infinity.
This divergence must be regulated through an infrared cutoff, which we shall introduce in the form of a parameter $m$ of the order of the pion mass.
The CGC result for the integrated variance is, to leading logarithmic accuracy~\cite{Giacalone:2019kgg}:
\begin{equation}
 \label{xicgc}
 \xi({\bf r})\approx\frac{2\pi(N_c^2-1)}{g^4N_c^2}Q_B^4 Q_A^2\ln\left(1+\frac{Q_A^2}{m^2}\right)+(A\leftrightarrow B),
\end{equation}
where we have regulated the argument of the logarithm to ensure that the variance is always positive~\cite{Giacalone:2019kgg}.

We now construct an event-by-event prescription for $\rho({\bf r})$ which incorporates the CGC results, Eqs.~(\ref{1pointcgc}) and (\ref{xicgc}).
We assume that, in a given event, each nucleus contains \textit{sources}~\cite{Bhalerao:2011bp} located at transverse coordinates ${\bf s}_j$, where $j$ labels the source, and that $\rho({\bf r})$ after the collision is the superposition of the contributions of individual sources coming from both nuclei: 
\begin{equation}
  \label{density}
  \rho({\bf r})=Q_B^2\sum_{j\in A} \Delta_A({\bf r}-{\bf s}_j)+
  Q_A^2\sum_{j\in B} \Delta_B({\bf r}-{\bf s}_j)
\end{equation}
where $\Delta_{A/B}$ is the profile of a source from nucleus $A/B$, to be specified below.
We assume that the positions ${\bf s}_j$ are independent random variables, and we denote by $n_{A/B}$ the density of sources in nucleus $A/B$.
Note that the energy density profile defined by Eq.~(\ref{density}) is a sum of two terms, symmetric under the exchange of $A$ with $B$.
Each term is the product of a random function involving one nucleus by the saturation momentum squared of the other nucleus.
This decomposition is illustrated in Fig.~\ref{fig:1}. 

The one- and two-point functions of the energy density defined by Eq.~(\ref{density}) are given by:
\begin{align}
  \label{12point}
  \nonumber \langle\rho({\bf r})\rangle&=Q_B^2n_A\int_{{\bf s}}\Delta_A({\bf r}-{\bf s})+(A\leftrightarrow B) ,\\
 S({\bf r}_1,{\bf r}_2)&=Q_B^4n_A\int_{{\bf s}}\Delta_A({\bf r}_1-{\bf s})\Delta_A({\bf r}_2-{\bf s}) + (A\leftrightarrow B).
\end{align}
We now determine $n_A$ and $\Delta_A({\bf r})$ such that one recovers the CGC results. 
By taking the ratio of the first terms in the right-hand side of these equations, one eliminates $n_A$. 
Thus, the ratio of the two-point function and the one-point function directly determines the source profile $\Delta_A({\bf r})$.\footnote{Interestingly enough, the ultra-violet divergences which plague the calculation of the energy density at proper time $\tau=0^+$ cancel when computing the ratio $S/\langle\rho\rangle$ (see Eqs.~(3.22) and (4.50) of Ref.~\cite{Albacete:2018bbv}). Hence the source profile, $\Delta_A({\bf r})$, at $\tau=0^+$ turns out to be a more robust quantity than the 1-point and the 2-point functions individually.}
As argued above, the most important quantity for phenomenology is the integrated variance (\ref{defxi}): 
\begin{equation}
\label{ximodel}
\xi({\bf r})=Q_B^4n_A\left[\int_{{\bf s}}\Delta_A({\bf r}-{\bf s})\right]^2+(A\leftrightarrow B).
\end{equation}
Substituting the expressions from the CGC, Eqs.~(\ref{1pointcgc}) and (\ref{xicgc}), and assuming that the two terms in the right-hand side of the first line of Eq.~(\ref{12point}) contribute symmetrically ($\frac{1}{2}$ each), one obtains:
\begin{equation}
  \label{sumrule}
  \int_{{\bf s}}\Delta_A({\bf r}-{\bf s})=
  \frac{8\pi}{g^2N_c}\ln\left(1+\frac{Q_A^2}{m^2}\right). 
\end{equation}
Choosing the following source profile: 
\begin{equation}
\label{sourceprofile}
  \Delta_{A}({\bf r})
  =
  \left\{
  \begin{aligned}
    \frac{8}{g^2N_c}\frac{1}{|{\bf r}|^2+Q_{A}^{-2}}, &\qquad& |{\bf r}|<1/m, \\
    0,  &&  |{\bf r}|>1/m.
  \end{aligned}
  \right.
\end{equation}
one recovers Eq.~(\ref{sumrule}), and also the $1/|{\bf r_1}-{\bf r_2}|^2$ decrease of the two-point function at large distances.\footnote{The two-point function defined by the second line of Eq.~(\ref{12point}) involves the convolution of $\Delta_{A}({\bf r})$ with itself, which is proportional to $1/|{\bf r_1}-{\bf r_2}|^2$, up to a slowly-varying logarithm.}

We now discuss the physical interpretation of Eq.~(\ref{sourceprofile}).
The large-distance behavior of $\Delta_A({\bf r})$ is that of a Coulomb field in two dimensions.
The electric field decreases like $1/r$, and its energy density like $1/r^2$.
Note that $\Delta_{A}({\bf r})$ goes to a finite value for $r\to 0$, while it would diverge for a pointlike charge.
The physical interpretation is that the charge is spread over a distance $\sim 1/Q_A$.
Now, the number of elementary charges contained in an area of this size is of order $1/g^2$, which explains the corresponding factor in Eq.~(\ref{sourceprofile}). 
Finally, each source profile in Eq.~(\ref{density}) is multiplied by the saturation momentum of the incident nucleus. 
This is related to the fact that, during the collision process, the 
(initially transverse) chromo-electric and chromo-magnetic fields 
acquire longitudinal components~\cite{Lappi:2006fp}. This is a genuine non-Abelian effect, that one may view as resulting from the fusion of two gluons (one from each nucleus). Thus, the longitudinal fields produced during 
the collision must be proportional to the color charges of both nuclei, 
hence the factor $Q_B^2$ ($Q_A^2$) in the first (second) 
term of Eq.~(\ref{density}).

The density of sources, $n_A$, is eventually obtained from the first line of Eq.~(\ref{12point}). Substituting Eqs.~(\ref{1pointcgc}) and (\ref{sumrule}), we obtain: 
\begin{equation}
  \label{densityofsources}
n_{A}=\frac{N_c^2-1}{32\pi}\frac{Q_{A}^2}{\ln\left(1+\frac{Q_{A}^2}{m^2}\right)}.
\end{equation}
It is naturally proportional to the number of gluon colors $N_c^2-1$ since the CGC is an effective theory of small-$x$ gluons~\cite{Gelis:2010nm}.
Up to the logarithmic correction, it is also proportional to $Q_{A}^2$, which is the typical behavior for the density of partons in the McLerran-Venugopalan (MV) model~\cite{McLerran:1993ni}. 

So far we have treated nuclei as infinite and uniform, with constant saturation momenta $Q_A$ and $Q_B$. 
In order to apply this framework to phenomenology, one needs to take into account the finite size of the nucleus, or, equivalently, the dependence of $Q_A$ and $Q_B$ on transverse coordinates.
This can be done unambiguously if $Q_A$ and $Q_B$ vary over a scale much larger than other scales in the system, in particular the infrared scale $1/m$.
The straightforward generalization of Eq.~(\ref{density}), taking into account the variation of $Q_{A/B}$, is: 
\begin{equation}
  \label{densitylocal}
  \rho({\bf r})=  \sum_{j\in A}Q_B^2({\bf s}_j) \Delta_A({\bf r}-{\bf s}_j)
+(A\leftrightarrow B). 
\end{equation}
Similarly, one takes into account the variation of $Q_{A/B}$ when sampling the sources according to Eq.~(\ref{densityofsources}). 
In Eq.~(\ref{sourceprofile}), one simply replaces $Q_{A/B}$ with its value at the center of the source.  

In summary, the present model provides a transparent physical picture of the Glasma~\cite{Lappi:2006fp} energy density profile as a superposition of contributions coming from collisions between localized charges and a uniform nucleus.
Note that fluctuations in this picture solely arise from the positions of the charges, which are sampled randomly over the area of the nucleus.
This is reminiscent of the Monte Carlo Glauber model, where fluctuations originate from positions of nucleons within the nucleus~\cite{Miller:2007ri}. 
The collision process itself, on the other hand, is {it deterministic\/}.
The physical interpretation is that the Glasma dynamics is governed by {\it classical\/} field equations. 

The model is particularly suitable for a Monte Carlo implementation, which we shall use in the remainder of this paper, and which we name \texttt{magma}. 
Let us move, then, to a phenomenological application.

In this article, as in Ref.~\cite{Giacalone:2019kgg}, we assume that $Q_{A/B}^2$ is proportional to the integral of the nuclear density over the longitudinal coordinate, which is usually denoted by $T_{A/B}$~\cite{Miller:2007ri}.
Thus, the only free parameter in our calculation is the proportionality constant or, equivalently, the value of the saturation momentum at the center of the nucleus, which we denote by $Q_{s0}$. 
Note that our approach differs from that of the IP-Glasma model~\cite{Schenke:2012wb}, in which one samples the positions of nucleons within each nucleus, and then evaluates locally the saturation momentum depending on these positions.
We consider instead that fluctuations associated with the wavefunctions of incoming nuclei are already effectively taken care of by sampling the sources ${\bf s}_j$ within each nucleus. 
\begin{figure*}[t]
    \centering
    \includegraphics[width=\linewidth]{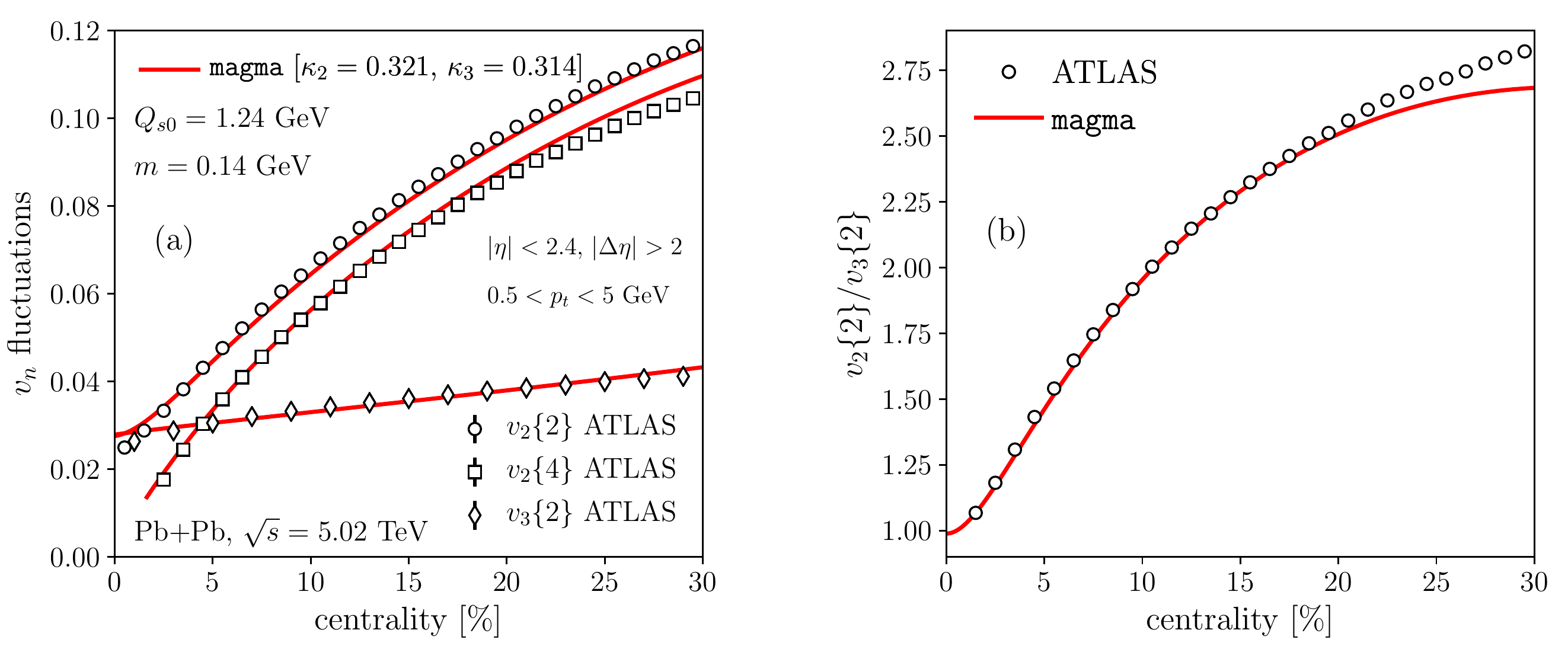}
    \caption{
      Symbols: Experimental data on $v_2$ and $v_3$, as function of centrality percentile, measured by the ATLAS Collaboration~\cite{Aaboud:2019sma} in 5.02 TeV Pb+Pb collisions.
      Lines: results from the magma calculation.
Panel (a) shows $v_2\{2\}$, $v_2\{4\}$ and $v_3\{2\}$, while panel (b) displays the ratio $v_2\{2\}/v_3\{2\}$.}
    \label{fig:2}
\end{figure*}

Figure~\ref{fig:1} presents a Monte Carlo realization of our model for a central Pb+Pb collision. 
The points on the left plot represent the positions of the sources in each nucleus, sampled according to the density (\ref{densityofsources}).
Note that there are sources lying outside the nucleus.
The reason is that for small $Q_{A}$, the density of sources (\ref{densityofsources}) does not vanish, but goes to a constant proportional to $m^2$.
However, these outliers are physically irrelevant, as they give a negligible contribution to the energy density, as explicitly observed on the right panel of Fig.~\ref{fig:1}, where most of the energy is within the nuclear radius. 
Note that void areas appear, even close to the center.
These void areas are filled during the early pre-equilibrium evolution of the system, before the hydrodynamics applies~\cite{Kurkela:2018wud}. 
Note that, in the case of a central collision as in Fig.~\ref{fig:1}, the contribution of each source to the energy density only depends on its distance to the center. 

We now apply this model to the description of anisotropic flow in Pb+Pb collisions.
The coefficient of anisotropic flow, $v_n$, is defined as the $n$-th Fourier harmonic of the azimuthal distribution of outgoing particles~\cite{Luzum:2011mm}. 
The largest harmonics in the spectrum are elliptic flow, $v_2$, and triangular flow, $v_3$. 
Hydrodynamic simulations show that, in a given class of impact parameter, $v_n$ is to a good approximation~\cite{Gardim:2011xv,Niemi:2012aj,Gardim:2014tya} linearly correlated with the initial anisotropy of the density profile $\rho({\bf r})$ in harmonic $n$, which is given by~\cite{Teaney:2010vd}:
\begin{equation}
  \label{defepsn}
\varepsilon_n\equiv \frac{\left|\int_{z}({z}-{z}_0)^n\rho({z})\right|}{\int_{z}|{z}-{z}_0|^n\rho({z})},
\end{equation}
where we have used the complex notation ${z}\equiv x+i y$, and $z_0$ is the center of energy:
\begin{equation}
  \label{defs0}
  z_0\equiv \frac{\int_{z} z\rho({z})}{\int_{z} \rho({z})}. 
\end{equation}
Therefore, in a narrow bin of collision centrality we can use $v_n\simeq \kappa_n\varepsilon_n$, 
where $\kappa_n$ is a positive response coefficient, whose centrality dependence can be neglected if one focuses on central collisions~\cite{Noronha-Hostler:2015dbi}.

Anisotropic flow is not measured on an event-by-event basis.
Quantities accessible experimentally are moments, or cumulants of the distribution of $v_n$.
The lowest order cumulants are~\cite{Borghini:2001vi}:
\begin{eqnarray}
v_n\{2\}&\equiv&\sqrt{\langle |v_n|^2\rangle}\cr
v_n\{4\}&\equiv&\left(2\langle |v_n|^2\rangle^2-\langle |v_n|^4\rangle\right)^{1/4},
\end{eqnarray}
where angular brackets denote an average value over many events in a narrow centrality class. 
The most accurately measured cumulants are $v_2\{2\}$, $v_2\{4\}$, $v_3\{2\}$~\cite{Acharya:2018lmh,Chatrchyan:2013kba,Aaboud:2019sma}.
Their centrality dependence in $5.02$~TeV Pb+Pb collisions is displayed in Fig.~\ref{fig:2} (a). 
Linear response implies that they are proportional to the corresponding cumulants of the initial anisotropy $\varepsilon_n$:
\begin{eqnarray}
  \label{rescaling}
v_2\{2\}&=&\kappa_2\varepsilon_2\{2\},\cr
v_2\{4\}&=&\kappa_2\varepsilon_2\{4\},\cr
v_3\{2\}&=&\kappa_3\varepsilon_3\{2\}.
\end{eqnarray}
Our Monte Carlo calculation goes as follows: We sample the position of the sources in each nucleus, we calculate $\varepsilon_n$ for each event by inserting Eq.~(\ref{density}) into Eqs.~(\ref{defs0}) and (\ref{defepsn}), and we evaluate the integrals analytically\footnote{Since the integral in the denominator of $\varepsilon_3$ cannot be evaluated analytically, we make the approximation that the relative fluctuations of $\langle r^3\rangle$ are identical to the relative fluctuations of $\langle r^2\rangle^{3/2}$ at every centrality.} for different impact parameters.
We treat $\kappa_2$, $\kappa_3$, and $Q_{s0}$ as free parameters, which we adjust to data. 
The values of $Q_{s0}$ and $m$ are the same as in Ref.~\cite{Giacalone:2019kgg},\footnote{Changing the cutoff amounts to renormalizing the value of $Q_{s0}$ (Appendix B of Ref.~\cite{Giacalone:2019kgg}) but does not change the physical quantities in the chosen centrality range.}  where the same quantities were evaluated directly as a function of the 1-point and 2-point functions within a small-fluctuation approximation~\cite{Blaizot:2014nia}.
The response coefficients $\kappa_2$ and $\kappa_3$ required to match data are by a few percent larger than in Ref.~\cite{Giacalone:2019kgg}. 
The main reason for this small difference is that our new calculation consistently takes into account the spatial extension of individual sources. 
Our results, displayed as lines in Fig.~\ref{fig:2}, are in very good agreement with experimental data. 
In particular, the centrality dependence of the ratio $v_2\{2\}/v_3\{2\}$, which is typically missed by hydrodynamic calculations~\cite{Alba:2017hhe}, is naturally reproduced. 

In conclusion, we have proposed a simple model of event-by-event fluctuations in heavy-ion collisions, where the energy density in each event has a simple analytic form given by Eq.~(\ref{density}). 
Our model reproduces the features of the one-point and two-point functions computed in the Glasma approach, to leading logarithmic accuracy.
Equations~(\ref{density}) and (\ref{sourceprofile}) give an intuitive picture of the initial density in a heavy-ion collison, as the sum of contributions from two-dimensional Coulomb fields of charges spread over a distance of order $1/Q_s$. 
Implementing this effective description through Monte Carlo calculations is straightforward and numerically fast. 
It provides a robust initial condition for pre-equilibrium and hydrodynamic studies, motivated by QCD. 

\section{acknowledgements}
J.Y.O thank Jean-Paul Blaizot and Matthew Luzum for discussions.
P.G-R. acknowledges financial support from the `La Caixa' Banking Foundation.
The work of CM was supported in part by the Agence Nationale de la Recherche under the project ANR-16-CE31-0019-02. FGâ€™s work was supported by the Agence Nationale de la Recherche through the
project 11-BS04-015-01.

\end{document}